\documentclass[aip,jcp,twocolumn,superscriptaddress]{revtex4}
\usepackage{graphicx}
\usepackage{amsmath}
\usepackage{amssymb}
\usepackage{pifont}
\usepackage{color}

\begin{document}

\title{Electrostatic interaction of neutral semi-permeable membranes}

\author{Olga I.~Vinogradova}

\affiliation{A.N.~Frumkin Institute of Physical
Chemistry and Electrochemistry, Russian Academy of Sciences, 31
Leninsky Prospect, 119991 Moscow, Russia}
\affiliation{Faculty of Physics, M. V. Lomonosov Moscow State University, 119991 Moscow, Russia}

\author{Lyderic~Bocquet}
\affiliation{LPMCN, University of Lyon, Lyon, France}

\author{Artem N.~Bogdanov}
\affiliation{A.N.~Frumkin Institute of Physical
Chemistry and Electrochemistry, Russian Academy of Sciences, 31
Leninsky Prospect, 119991 Moscow, Russia}
\author{Roumen~Tsekov}
\affiliation{Faculty of Chemistry, University of Sofia, 1164 Sofia, Bulgaria}
\affiliation{DWI,  RWTH Aachen, Forckenbeckstr. 50, 52056 Aachen, Germany}

\author{Vladimir~Lobaskin}
\affiliation{School of Physics and Complex and Adaptive Systems
Laboratory, University College Dublin, Belfield, Dublin 4,
Ireland}

\date{\today}
\begin{abstract}
We consider an osmotic equilibrium between bulk solutions of
polyelectrolyte bounded by semipermeable membranes
and separated by a thin film of salt-free liquid. Although the membranes are neutral, the counter-ions of the polyelectrolyte molecules permeate into the gap and lead to a steric charge separation. This gives rise to a distance-dependent membrane potential, which translates into a repulsive electrostatic disjoining pressure. From the solution of the non-linear Poisson-Boltzmann equation we obtain the distribution of the
potential and of ions. We then derive an explicit formula for the pressure
exerted on the membranes and show that it deviates from the classical van't Hoff expression for the osmotic pressure.
 This difference is interpreted in terms of a repulsive electrostatic disjoining pressure originating from the overlap of counterion clouds inside the gap.
  We also develop a
simplified theory based on a linearized Poisson-Boltzmann
approach. A comparison with simulation of a primitive model for the electrolyte
is provided and does confirm the validity of the theoretical predictions
Beyond the fundamental
result that the neutral surfaces can repel, this mechanism not
only helps to control the adhesion and long-range interactions of
living cells, bacteria, and vesicles, but also allows us to argue
that electrostatic interactions should play enormous role in
determining behavior and functions of systems bounded by
semipermeable membranes.

\end{abstract}
 \maketitle
\section{Introduction}

It is hard to overestimate the role semipermeable membranes and
osmotic equilibria associated with them play in our everyday life.
The best known examples are the natural biological membranes, which are highly impermeable to ions due to their phospholipid bilayer structure, but become semi-permeable when ion channels are open. Such lipid membranes
with channel proteins surround all biological (eukaryotic and
prokaryotic) cells~\cite{sheeler.p:1987}. Many synthetic membranes
used in electrochemical fuel cells~\cite{winter.m:2004} and
dialysis~\cite{maher.jf:1977} take advantage of the
semipermeability of their materials. The same concerns various
types of synthetic vesicles~\cite{menger.fm:1998,vinogradova.oi:2006,vinogradova.oi:2004b},
the promising
gene and drug carriers, as well as other systems used to mimic
biological objects. Being in contact with ionic solutions,
such as gels, polyelectrolytes (including DNA, proteins,
dendrimers), micelles, or colloids, a semipermeable membrane
maintains an unequal distribution of ionic solute concentrations,
which leads to an ion density gradient across the membrane, determines its
actual potential, and generates an osmotic pressure difference
that the membrane has to sustain. This situation is traditionally
referred to as a Donnan equilibrium~\cite{donnan.fg:1924,donnan.fg:1995}.

\begin{figure}
\includegraphics[width=4.5cm,clip]{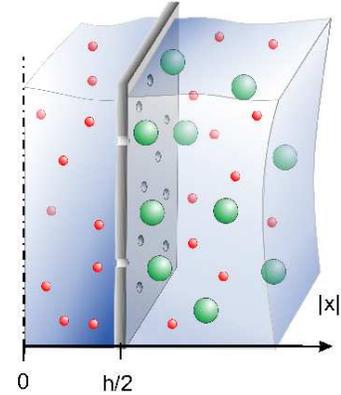}
\caption{An electrolyte solution in equilibrium with a thin
liquid film of thickness $h$ bounded by neutral semipermeable membranes.
They are permeable for small counterions only (here anions), while the larger solute (here cations) are bounded to the region $\vert x\vert > h/2$. The membranes have to
support an excess osmotic pressure $\Delta p (h)$ which is the sum
of the bulk Donnan pressure and an electrostatic disjoining
pressure $\Pi (h)$ in the interlayer.} \label{fig:capsule}
\end{figure}

A great deal of research has been devoted to understanding the
Donnan equilibria and the pressure exerted on semipermeable
membranes. Although the problem was treated at various levels of
sophistication, traditionally the reservoirs separated by a
semipermeable wall are assumed to be sufficiently large, so
that both solutions contain a phase with the bulk properties. Our
paper deals with another situation, when two ionic electrolyte
solutions consisting of large ions and small ions (referred below
to as counter-ions) are in equilibrium with a thin film bounded by
the semipermeable membranes. As some mobile
counter-ions will inevitably escape from electrolyte
solutions, at some separation their clouds will begin to overlap and
give rise the effect similar to the celebrated Derjaguin
disjoining pressure~\cite{derjaguin.b:1941} and will change
dramatically the Donnan equilibrium~\cite{donnan.fg:1924} in the
system. Some experimental observations~\cite{kim.bs:2007} support
this idea. However, to our surprise, such a scenario, which represents
enormous interests for many biological and materials science
problems associated with the membrane adhesion and long-range
interactions~\cite{razatos.a:1998,jucker.ba:1996,gingell.d:1977},
and are also relevant to modern micro- and nanofluidics~\cite{schoch.rb:2008,jong.j:2006},
has never been addressed before~\cite{note2}.

In our previous papers we made an attempt to calculate
an excess pressure on a semipermeable shell in contact with an
inner~\cite{tsekov.r:2006} or outer~\cite{tsekov.r:2008} solution of polyions.
Since a non-linear Poisson-Bolzmann (NLPB) equation cannot be solved analytically
for spherical geometry~\cite{andelman.d:1995}, its linearized version (LPB) has been used.
Our present paper solves a pressure problem for a flat geometry of two interacting
semipermeable membranes. We first solve semi-analytically a NLPB equation to evaluate
the distribution of electrostatic potential in the system. We then derive an explicit expression
for a pressure on the membranes and a disjoining pressure in the gap between them.
Our mean-field approach is verified for monovalent salts by molecular dynamics (MD)
simulations. Simulation data fully support our theory.

Our paper is organized as follows: In Section II some general consideration concerning a theoretical
description of an interaction between two semi-permeable membranes are presented. Here we also
describe a simplified linearized version of the theory. Section III contain a description of our
MD simulation approach. In Section IV simulation results are presented to validate the predictions of the theory.

\section{Theory}

The geometry of the system under consideration is shown in Fig.~\ref{fig:capsule}.
A semipermeable membrane is in contact with a solution of
polyelectrolyte composed of cations with an effective charge $Z$
and concentration $C$, anions with charge $z$ and concentration $c$. We assume here that
the polyelectrolyte (here cations) cannot permeate through the semi-permeable membrane,
while their counter-ions (here anions) are free to pass through it.
The membrane is at distance $h$ from another membrane.

To make the formulas as transparent as possible we keep our
analysis at the mean-field level by using the Poisson-Boltzmann approach.
This means that we treat ions as point-like and neglect their correlations. In particular, while the results of the Poisson-Boltzmann theory to be discussed below can be computed for any valence $Z$ of the macromolecules, correlations between macro-ions should be taken into account in the limit of large charges $Z$ in order to obtain quantitative predictions. However, based on earlier results~\cite{tsekov.r:2008}, one does not expect the main physical picture to be altered in this case, and we leave the study of this regime for a future work.

We consider a hypothetical case of an infinitesimally thin and rigid membrane.
We further assume that both membranes are neutral, and axis $x$ is directed normally
to the surfaces with $x=0$ at the midplane of the gap. The membranes are located at $|x|=h/2$.
Our description thus essentially follows that of the classical non-linear Poisson-Boltzmann
theory, except for the fact that there is no charge {\it per se} on the membranes and the membrane
surface potential builds up self-consistently: accordingly the distribution of the charged species
is a consequence of the semipermeable character of the membrane, which leads to a steric charge separation.

\subsection{Non-linear theory}
\subsubsection{Potential}

We first introduce the dimensionless electrostatic potentials
\begin{equation}
\phi_{i,o}={z e\varphi_{i,o}\over k_BT}
\end{equation}
with the index $\{i,o\}$ standing for ``in'' ($\vert x\vert < h/2$) and ``out'' ($\vert x\vert > h/2$) of the confined slab.

The non-linear Poisson-Boltzmann (NLPB) equation then reads
\begin{eqnarray}
& \Delta\phi_o&= - \kappa_i^2 \left( e^{-\phi_o}-e^{-\tilde Z \phi_o} \right) \label{NLPB-1}
\\
& \Delta\phi_i&= - \kappa_i^2\, e^{-\phi_i}
\label{NLPB-2}
\end{eqnarray}
where the inner inverse Debye screening length, $\kappa_i$, is defined as $\kappa_i^2=4\pi \ell_B c_\infty$ with $\ell_B=z^2e^2/(4\pi\epsilon\epsilon_0k_BT)$ the Bjerrum
length, $\tilde Z=Z/z$ ($<0$) is the valence ratio of large and small ions, and $ c_{\infty}$ is the concentration of small ions far from the membrane. The outer inverse Debye screening length, $\kappa_o$, can be defined as $\kappa_{o}^2 = 4\pi \ell_B (\tilde Z^2 C_{\infty}+c_\infty)$, where $ C_{\infty}$ is the concentration of large ions far from the membrane. Obviously, it represents the inverse Debye length of the bulk electrolyte solution. Since the electroneutrality condition $Z
C_{\infty} + z c_{\infty} = 0$ is employed, $\kappa_o = \kappa_{i} \sqrt{1 - \tilde Z}$. We  stress, however, that for this
particular problem, the main reference length scale that determines the behavior of the system is $\kappa_{i}^{-1}$,
and this is reflected in the analysis below.

The first integration gives a differential equation for $\phi_o$
\begin{equation}
{1\over 2\kappa_i^2} \left({\partial \phi_o\over \partial x}\right)^2=e^{-\phi_o}-{1\over\tilde Z} e^{-\tilde Z \phi_o} + A_0
\label{dphi-o}
\end{equation}
where the integration constant $A_0$ is determined by the boundary conditions at infinity: $\phi_o\rightarrow 0$ and
$\partial_x\phi_o \rightarrow 0$. Imposing them on (\ref{dphi-o}) yields  $A_0={1\over\tilde Z} -1$.
Thus, the outer solution $\phi_o$ can be obtained in terms of the membrane ``surface'' potential $\phi_s=\phi(h/2)$
\begin{eqnarray}
&&\int_{\phi_s}^{\phi_o}{d\phi \over \sqrt{2\left( \exp[-\phi]-1- {1\over\tilde Z} \left( \exp(-\tilde Z \phi)-1\right)\right)}}\nonumber\\&=&- \kappa_i\left(x-{h\over2}\right)
\label{xout}
\end{eqnarray}

For the inner compartment, $\phi_i$, the first integration of the NLPB equation yields
\begin{equation}
{1\over 2\kappa_i^2} \left({\partial \phi_i\over \partial x}\right)^2=e^{-\phi_i}-e^{-\phi_m}
\label{dphi-i}
\end{equation}
Here we used $\partial_x\phi(x=0)=0$ implied by symmetry. This leads to
\begin{equation}
\int_{\phi_m}^{\phi_i}{d\phi \over \sqrt{2\left( \exp[-\phi]-\exp[-\phi_m] \right)}}=- \kappa_i\,x
\label{xin}
\end{equation}
with $\phi_m$ the (dimensionless) potential at the center of the film between membranes.
This integral can be evaluated exactly as
\begin{equation}
2 \arctan\left[\sqrt{\exp[-(\phi_i-\phi_m)]-1}\right] \exp\left[{\phi_m\over 2}\right]=\sqrt{2}Ê\kappa_i x,
\end{equation}
leading to the Gouy-type expression
\begin{equation}
\phi_i(x)=\phi_m+\ln\left[ \cos^2\left({\sqrt{2}\over 2} e^{-\phi_m/2} \kappa_i x\right)\right].
\label{phin-ex}
\end{equation}

The continuity of the electric field at the membrane surface, Eqs. (\ref{dphi-o})-(\ref{dphi-i}) (no surface charge) leads to
\begin{equation}
e^{-\phi_s}-{1\over\tilde Z} e^{-\tilde Z \phi_s} +{1\over\tilde Z} -1=e^{-\phi_s}-e^{-\phi_m}
\end{equation}

Altogether, the membrane potential $\phi_s$ and mid-plane potential $\phi_m$ are thus given
by the self-consistent equations
\begin{equation}
\phi_s=\phi_m+\ln\left[ \cos^2\left({\sqrt{2}\over 2} e^{-\phi_m/2} \kappa_i {h\over 2}\right)\right]
\label{SC-1}
\end{equation}
and
\begin{equation}
\phi_m=- \ln\left[ 1 +{1\over\tilde Z} \left(e^{-\tilde Z \phi_s}-1\right)\right]
\label{SC-2}
\end{equation}

In the general case, the derived equations should be solved numerically while the
in the limits of large and small $\kappa_{i} h$ we can also find the asymptotic analytical expressions.
In the thick gap limit, $\kappa_i h \gg 1$, the midplane potential diverges, $
\phi_m\rightarrow \infty$, and the equation for $\phi_s$, Eq. (\ref{SC-2}), can be simplified to give
\begin{equation}
\phi_s \simeq - {1\over \tilde Z}\ln(1-\tilde Z)
\label{phis-asymptotic}
\end{equation}
This value represents the bulk Donnan potential.

Similarly, the asymptotic behavior for $\phi_m$ can be obtained
from Eq. (\ref{SC-1}). Since $\phi_s$ is bounded by a constant,
the condition $\phi_m\gg 1$ as $\kappa_i h\gg 1$
imposes that ${\sqrt{2}\over 2} e^{-\phi_m/2} \kappa_i {h\over 2} \simeq \pi/2$.
This leads to
\begin{equation}
\phi_m\simeq 2 \ln\left[ {\sqrt{2}\over 2\pi}\kappa_i h\right]
\label{phim-asymptotic}
\end{equation}

In the thin gap limit, $\kappa_i h \ll 1$, where the inner ionic
clouds strongly overlap, both $\phi_m$ and $\phi_s$
vanish. Such a situation would be realistic for very
dilute polyelectrolyte solutions and/or very thin gap. One can
easily verify that $\phi_m\simeq \phi_s \propto \kappa_i h$.

Thus, the
convergence of two semipermeable membranes is necessarily
accompanied by the decrease in the absolute value of their
potential. The idea that the constant potential condition is not
appropriate for fully permeable charged membranes has been
suggested before~\cite{gingell.d:1967,ninham.bw:1971}. Now we have
shown that the potential of neutral semipermeable surfaces
should inevitably change and can even vanish as a result of their
approach. Such a finding might be especially important for
biomembranes, where an alteration of a surface potential can lead
to a characteristic biological response.

\subsubsection{Osmotic and disjoining pressure}
The force balance in each part of the membrane (in and out) can be written
\begin{equation}
-\nabla p + \rho_c E =0
\end{equation}
with $\rho_c$ the charge density and $E=-\partial_x \phi$ the local electric field.
Using the Boltzmann expressions for the charge densities in terms of the local electrostatic
potentials allows to integrate this equation once.

In midspace between the membranes ($\vert x \vert <h/2$), this leads to
\begin{equation}
p_{i}(x)=k_BT c(x) + p_0
\label{pin}
\end{equation}
with $c(x)=c_\infty \exp[-\phi(x)]$ the counter-ion concentration
and $p_0$ a constant.

In the outer space ($\vert x \vert <h/2$) one gets
\begin{equation}
p_{o}(x)=k_BT c(x) + k_BT C(x) + p_L
\label{pout}
\end{equation}
with $C(x)=C_\infty \exp[-\tilde Z \phi(x)]$ the concentration of large ions and $p_L$ the
pressure of pure solvent.

At the membrane, there is a pressure drop due to the repulsion force acting on the
polyelectrolyte by the membrane (and proportional to the
difference of polyelectrolyte concentration on the two sides of the membrane), {\it i.e.}:
\begin{equation}
p_{o}\left({h\over 2}^+\right)-p_{i}\left({h\over 2}^-\right)=k_BT C\left({h\over 2}^+\right)
\label{DP}
\end{equation}
This imposes $p_0=p_L$, i.e. the solvent pressure, as expected.

The force acting on the membrane (osmotic pressure) can be found from the Maxwell tensor
$\mathbb{T}=(P+{\epsilon\over 2} E^2) \mathbb{I}-\epsilon E\bigotimes E$.
Using $\nabla\cdot\mathbb{T}=0$, we find the force per unit surface on the membrane as
$\Delta p= \mathbb{T}(x=0)- \mathbb{T}(x=\infty)$,
\begin{eqnarray}
&\Delta p&= \mathbb{T}(x=0)- \mathbb{T}(x=\infty)\nonumber\\
&& =k_BT c_\infty (1-e^{-\phi_m})+k_BT C_\infty
\label{finalP}
\end{eqnarray}
Note that by using Eq. (\ref{SC-2}) one can demonstrate
that expression given in Eq. (\ref{finalP}) is fully equivalent to
Eq. (\ref{DP}). Therefore, since the disjoining pressure, $\Pi$, is defined
via
\begin{equation}\label{def_Pi}
    \Delta p= k_BT(c_\infty+C_\infty)-\Pi,
\end{equation}
it can be expressed through the
concentration (or potential) at the mid-plane as
\begin{equation}
\Pi=k_BT c_m=k_BT c_\infty e^{-\phi_m}
\label{finalDP}
\end{equation}
In other words, the whole effect can be expressed through the osmotic pressure of small ions
in the mid-plane of the gap where the electric field vanishes. Note that the similar physical
interpretation of the disjoining pressure between flat solid
surfaces was given long ago in the famous work by
Langmuir~\cite{langmuir.i:1938}.
Note that the disjoining pressure is always positive indicating an
electrostatic repulsion between neutral semipermeable membranes
separating the similar electrolyte solutions.

By using the expression of the disjoining pressure, (\ref{finalDP}), we get in thick gap limit
\begin{equation}\label{tgl}
\Pi \simeq k_BT c_\infty \, {2 \pi^2\over \left( \kappa_i h\right)^2}
\end{equation}
{\it i.e.} $\Pi \propto h^{-2}$.
This is very similar to the famous Langmuir result, but here for an
{\it a priori} uncharged semi-permeable membrane.

In contrast, at small $\kappa_i h$
\begin{equation}\label{disprshort}
    \Pi \approx k_B T c_{\infty}
\end{equation}
This suggests that the osmotic pressure in the gap is dramatically
reduced compared to the value expected in the bulk:
\begin{equation}\label{poly_pr}
\Delta p=k_BT C_\infty = {1 \over 1-\tilde Z} \, p_{\rm{id}}
\end{equation}
with $p_{\rm{id}}= k_BT (C_\infty+c_\infty)$ the ``ideal'' (bulk) osmotic
pressure. Accordingly, $\Pi={-\tilde Z/(1-\tilde Z})\, p_{\rm{id}}$ in this limit.
This is one of the key results of our work.

\subsection{Linearized theory}

At low charge densities and low values of the electric potential, the description of the problem can be
simplified by linearization of the Poisson-Boltzmann approach (LPB). The
linear approximation for the local concentrations reads: $c_{i,o}(x)=c_{\infty} \left(1 - \phi_{i,o}(x)\right),$ and
$C_{o} (x) = C_{\infty} \left(1 - \tilde Z \phi_{o} (x)\right)$. Substituting them into
Eq.~\ref{NLPB-2}, we get its linearized version with the straightforward solutions
\begin{equation}\label{potential1}
  \phi_{o} = \phi_s \exp \left[\kappa_{o}\left(h/2-|x| \right) \right],
\end{equation}
\begin{equation}\label{potential2}
  \phi_{i} = 1 + \left(\phi_s - 1 \right)
  \frac{\cosh (\kappa_{i} x)}{\cosh\left(\kappa_{i}\displaystyle{h/2}\right)},
\end{equation}
The electroneutrality of the membrane allows us to deduce
\begin{equation}\label{membrane}
    \phi_s = \frac{\kappa_{i}}{\kappa_{i} + \kappa_{o}\coth \left(\kappa_{i}\displaystyle{h/2}\right)}
\end{equation}
\begin{equation}\label{wall}
    \phi_m =  1 - \frac{\kappa_{o}}{\kappa_{o}\cosh
    \left(\kappa_{i}\displaystyle{h/2}\right) + \kappa_{i}\sinh{\left(\kappa_{i}\displaystyle{h/2}\right)}}
\end{equation}

In the thin gap limit, the membrane
potential vanishes similarly to the non-linear case. For large gaps, we get
\begin{equation}\label{phis-asymptoticlin}
\phi_s \simeq \frac{\kappa_i}{\kappa_i + \kappa_o} = \frac{1}{1 +
\sqrt{1-\tilde Z}}
\end{equation}
The value of this bulk Donnan potential is different from predicted
by NLPB theory, Eq.(\ref{phis-asymptotic}).

We also note that the potential in the midplane
at large $\kappa_ih$ saturates, asymptotically approaching $1$ (in contrast to its
divergence in the NLPB theory, cf. Eq.(\ref{phim-asymptotic})).
This is an evidence of a failure of the linear theory in calculation of potentials
(and relevant ion profiles in the system).

Motivated by recent analysis~\cite{tsekov.r:2008,deserno.m:2002,dobnikar.j:2006} we then obtain the following expression for the pressure
\begin{equation}\label{po}
  p_{o} = p_L + k_B T (C_{\infty} + c_{\infty}) + \frac{\epsilon_0 \epsilon \kappa_o^2 \phi_o^2}{2},
\end{equation}
\begin{equation}\label{pi}
  p_{i} = p_L + k_B T c_m - z e c_{\infty} (\phi_i - \phi_m) + \frac{\epsilon_0 \epsilon \kappa_i^2 (\phi_i^2 -\phi_m^2)}{2}
\end{equation}
which contains quadratic terms to provide thermodynamic self-consistency of the linear theory.

Using these expressions, one can calculate the pressure difference
on both sides of the membrane to obtain~\cite{note3}
\begin{equation}\label{deltap}
    \triangle p=p_{\rm id} - k_B T c_m
\end{equation}
As one can easily see, all the electrostatic terms cancel and the
whole effect is expressed by the osmotic pressure of small ions
in the mid-plane of the gap. In other words, physically we have arrived to the
same result as in the NLPB case. Clearly, in the LPB case $c_m = c_i (x=0)$ is
very different, so that the disjoining pressure takes the form
\begin{equation}\label{disjpressure}
    \Pi= k_B T c_{\infty} \frac{\kappa_{o}}{\kappa_{o}\cosh
    \left(\kappa_{i}\displaystyle{h/2}\right) + \kappa_{i}\sinh{\left(\kappa_{i}\displaystyle{h/2}\right)}}
\end{equation}

For large
$\kappa_i h$, one can note some similarity to a repulsion of
solids~\cite{derjaguin.b:1941,parsegian.va:1972}, which exponentially
decays to zero as
\begin{equation}\label{disprlarge}
    \Pi \simeq k_B T c_{\infty} \frac{2 \kappa_o}{\kappa_o+\kappa_i}
    \exp \left(-\frac{\kappa_i h}{2} \right),
\end{equation}
which obviously differs from NLPB result, Eq.(\ref{tgl}).
For small $\kappa_i h$ we again get Eq.(\ref{disprshort}).

\section{Simulations}

The Langevin dynamics (MD) simulations were performed on the level
of the primitive model with explicit large and small ions using
the ESPResSo simulation package~\cite{Espresso}. We
constructed a 1D-periodic setup with two membranes fixed
perpendicular to the $x$-axis. The membranes were made impermeable
for cations, but ``invisible'' for anions
\cite{stukan.mr:2006}.

For an initial illustration of our approach we here use a
monovalent electrolyte, and ionic species were represented by
Lennard-Jones spheres with a central charge $Z = 1$ or $z = -1$.
We used the repulsive Lennard-Jones (RLJ) potential
with the cut-off distance $r_c=2^{1/6}\sigma$:
\begin{eqnarray}
U_{LJ}(r)=
\begin{cases}
4\epsilon\left[\left(\frac{\displaystyle\sigma}{\displaystyle r}\right)^{12}-\left(\frac{\displaystyle
\sigma}{\displaystyle r}\right)^6 +\frac{\displaystyle 1}{\displaystyle 4} \right], \quad r\leq r_c; \\
0,\quad r>r_c,
\end{cases}
\label{potlj_txt}
\end{eqnarray}
where $r$ is the distance between centers of two particles. The
energy parameter $\epsilon$ controls the strength of the
interaction, and its value was fixed to $\epsilon = 1.0 k_BT$.
The units of length and energy in all presented data were set by
$\sigma$ and $\epsilon$, respectively (LJ units). The bead
sizes were set as $\sigma_{pp}=\sigma_{cc}=\sigma_{pc}=1.0$.

The interaction of ions with the membrane was set by
\begin{eqnarray}
U_{LJ}(x)=
\begin{cases}
4\epsilon\left[\left(\frac{\displaystyle\sigma}{
\displaystyle x}\right)^{12}-\left(\frac{\displaystyle
\sigma}{\displaystyle x}\right)^6 -\left(\frac{
\displaystyle\sigma}{\displaystyle x_c}\right)^{12}+\left(\frac{
\displaystyle\sigma}{\displaystyle x_c}\right)^6\right],\\ \quad  x \leq x_c; \\
0,\quad x>x_c,
\end{cases}
\label{potlj_2}
\end{eqnarray}
with the cut-off distance $x_c=2^{1/6} \sigma$.

The solvent was treated as a homogeneous medium with a dielectric
permittivity set through the Bjerrum length. The electrostatic
interaction between the ionic species was modeled by the Coulomb
potential
\begin{equation}
U_{\rm Coul}(r_{ij})=k_BT\frac{\ell_B q_i q_j}{r_{ij}}
\label{colomb_txt}
\end{equation}
where $q_i=\pm1$. In all simulations we used $\ell_B =1$.

We modeled the systems in a rectangular cell with side $L$ ranging from
100 to 700 for different electrolyte concentrations. The number of
ions in the cell was varied from $N_p=500$ to 15000, and an
equivalent number of counterions was added. The number of ions in
each simulation was chosen to keep the ionic concentration in
center of the membrane fixed at $C_\infty = 0.03$ and to vary the value of
$\kappa_i h$ in a very large range, from 0.7 to 30.

\begin{figure}
\begin{center}
\includegraphics*[width=4.5cm,clip]{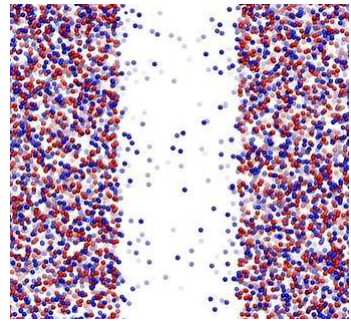}\\
\end{center}
{\caption{Side view of the film confined between two semi-permeable membranes. Cations are shown by blue spheres,
anions by the red spheres. One can see that anions leak out into the gap between the membranes.
 } \label{fig:mono_snap}}
\end{figure}
Three-dimensional periodic boundary conditions were used. For
Coulomb interactions we used the P3M algorithm with maximum
relative accuracy of $10^{-5}$. A snapshot of the system is
presented in Fig. \ref{fig:mono_snap}.

Pressure has been evaluated via integration of the LJ force of
cations, acting on the membrane walls
\begin{equation}\label{fx}
F(x) = 4 \varepsilon \left(\frac{12\sigma^{12}}{\left(x-\frac{h}{2}\right)^{13}}-\frac{6\sigma^{6}}
{\left(x-\frac{h}{2}\right)^{7}}\right)
\end{equation}
Using this force expression, we calculated the pressure as
\begin{equation}
p=\int_{h/2}^{h+2^{1/6} \sigma} C(x) F(x) dx.
\end{equation}

\section{Results and Discussion}

In this section we present results of MD computer simulations and some example calculations
based on the general NLPB theory as well as the analytical LPB results.

\begin{figure}
\begin{center}
\includegraphics*[width=8.5cm]{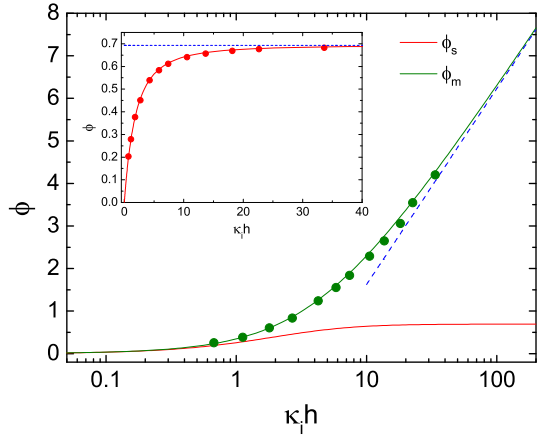}
\end{center}
{\caption{Midplane and surface potentials, $\phi_m$ (top curve), $\phi_s$ (bottom
curve), as a function of $\kappa_i h$. The dashed line is the asymptotic behavior
for $\phi_m$ in the large $\kappa_i h$ regime according to Eq.(\ref{phim-asymptotic}):
$\phi_m\approx 2\log
\left[\kappa_i h\sqrt{2}/2\pi\right]$. Inset shows zoom on the behavior of the surface potential
versus $\kappa_i h$. The dashed line here is the asymptotic behavior for $\phi_s$ in the large
$\kappa_i h$ regime according to Eq.(\ref{phis-asymptotic}): $\phi_s\approx -{1\over\tilde Z}\log(1-\tilde Z)$.
In these plots we use $\tilde Z=1$. }
\label{fig:potentials}}
\end{figure}

The distribution of the electrostatic potentials, $\phi_s$ and $\phi_m$, is shown in
Fig.~\ref{fig:potentials} versus $\kappa_i
h$ (symbols). Also included are the exact theoretical curve, calculated with
Eqs.(\ref{SC-1})-(\ref{SC-2}) (solid curves).
The agreement is excellent for all $\kappa_i
h$, even for very large values, confirming the validity of the mean-field approach for our system.
Asymptotic results are in agreement with the numerical calculations presented in
Fig. \ref{fig:potentials}. At $\kappa_i h \ll 1$ (strong overlap of an inner double layer), the membrane
potential vanishes. Such a situation would be realistic for very
dilute solutions and/or very thin gap. Another
asymptotic limit of large films and/or concentrated solutions,
$\kappa_ih \gg 1$ (no overlap of the inner ionic layers) gives
Eq.(\ref{phis-asymptotic}), which is fully supported by the simulation results.

\begin{figure}
\begin{center}
\includegraphics*[width=8.5cm]{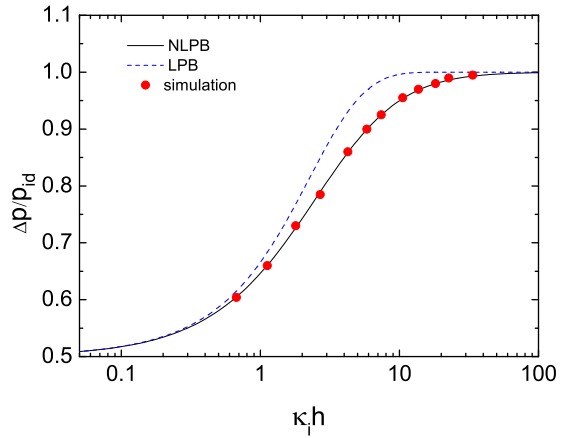}
\includegraphics*[width=8.5cm]{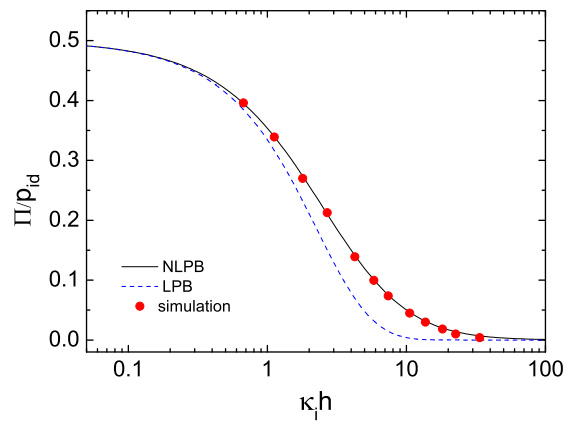}
\end{center}
{\caption{ Pressure acting on a membrane and a disjoining pressure
normalized by the ideal osmotic pressure, $p_{id}= k_BT(c_\infty+C_\infty)$.
Solid curves show NLPB predictions, dashed curve shows LPB theory predictions,
symbols present simulation results. }
\label{fig:Pi}}
\end{figure}

These pressure trends are illustrated in Fig.~\ref{fig:Pi}.  Simulations show that at large $\kappa_i h$
the pressure, $\Delta p/p_{\rm id}$ supported by the membrane, is
close to the osmotic pressure of the corresponding bulk solution
$\Delta p / p_{id} \approx 1$, and we deal with the standard bulk Donnan equilibrium. In this situation the disjoining pressure
is negligibly small. At smaller $\kappa_i h$ the pressure exerted on the semipermeable wall is much less than that
in the bulk, and at very small $\kappa_i h$ it approaches a constant, which is equal to the bulk osmotic
pressure of large ions, Eq.(\ref{poly_pr}). This is accompanied by an increase in the value of a disjoining
pressure in the gap. The results for the pressure and the disjoining pressure obtained in the NLPB theory and
simulations coincide confirming the validity of the mean-field approach for our system. Fig.~\ref{fig:Pi} also
includes the theoretical curves calculated within LPB theory. The agreement is quite good at very small and very
large $\kappa_i h$, but at intermediate values of $\kappa_i h$ there is
some discrepancy. The discrepancy is always in the direction of the pressure on membrane is larger than than
``measured'' in simulations and predicted by the NLPB theory. Correspondingly, the disjoining pressure is smaller.
Still, LPB and simulation pressures are in surprisingly good harmony, especially taking into account the simplicity of the
model and the complexity of the system. Obviously, the effects are somehow included in the quadratic term of the expression
for a pressure which provided a self-consistency of the LPB theory. These questions, however, deserve further investigation
and remain subject of a future work.

Finally, we note that all the results derived above hold
also for the case of a membrane placed at a distance $h/2$ from
the neutral wall suggesting that the tight adhesion of
semipermeable membrane to the neutral wall  is impossible within
our scenario. In general, for more complex systems, where the
membrane adhesion is controlled by the competition of several
effects, the physical mechanism we considered here should reduce
the attractive interactions by orders of magnitude similarly to
what was predicted for other types of electrostatic interactions
in the membrane systems~\cite{nardi.j:1998,sackmann.e:2002}. It
would be also worthwhile to emphasize that our derivation can
easily be modified for the situation where a thin film separates
the reservoirs with oppositely charged polyelectrolytes. In this
case, however, no attraction between semipermeable membranes takes
place as it would be tempting to expect. The point is that only
solution for such a configuration is $\phi=0$. Hence, all phases
are homogeneous and neutral. This result can, however, still be of
help when new synthetic delivery systems are designed. For
example, to avoid repulsion from the cell membrane a
semi-permeable neutral container should contain positively charged
molecules of drugs or proteins.

To summarize, we have examined theoretically the situation of an
interaction of two neutral semipermeable membranes separated by a
thin film. Our mechanism predicts an alteration of the membrane
potential during the approach, a decrease in osmotic pressure on
membranes when they are in a close proximity, and an electrostatic
repulsion between them. Our analysis also allows one to express a
disjoining pressure in the film through the osmotic pressure of
counter-ions in the midplane.

\section*{Acknowledements}

This research was partly supported by the RAS through its priority
program `Principles of basic studies of nanotechnologies and
nanomaterials' and FP7 project `BeyondEverest'.  Access to computational resources at the Center
for Parallel Computing at the M.V. Lomonosov Moscow State University
(`Lomonosov' and `Chebyshev' supercomputers) is gratefully
acknowledged.

\vfil

\bibliography{stukan3}

\begin{thebibliography}{30}
\bibitem[{not({\natexlab{b}})}]{note3}
\bibinfo{note}{We stress that it would be wrong in the linearized theory to
  evaluate pressure on the membrane as $k_B T C_o(h/2)$ as it has been
  discussed in previous
  work~\cite{deserno.m:2002,tsekov.r:2006,tsekov.r:2008}.}
\end{thebibliography}

\end{document}